# Self-regulated photoresponsive heterogeneous PNIPAM hydrogel actuators

Jingxuan Li[a‡], Jiaqi Miao[a‡], and Alan C. H. Tsang[a*]

[a] Department of Mechanical Engineering, The University of Hong Kong, Pokfulam, Hong Kong, China
[‡] These authors contributed equally to this work.
* Correspondence: alancht@hku.hk

**Abstract:** Self-regulated actuators harness material intelligence to enable complex deformations and dynamics, representing a significant advancement in automated soft robotics. However, investigations on self-regulated soft actuators, particularly those using simplified actuation modules such as a unidirectional light beam, remain limited. Here, we present a design paradigm for self-regulated actuators based on Poly(*N*-isopropylacrylamide) (PNIPAM) heterogeneous hydrogel, where self-regulated deformations are actuated by a fixed near-infrared laser. By utilizing the different responsiveness of PNIPAM hydrogels and those integrated with reduced graphene oxide (rGO), we develop three heterogeneous hydrogel configurations: up-down, side-by-side, and hybrid. These designs enable complex biomimetic deformations in soft hydrogel actuators, resembling a bending finger or a flexible industrial manipulator, all actuated by a single fixed laser source. These proposed heterogeneous designs and actuation strategies leverage material intelligence to create soft actuators with enhanced autonomy, paving the way for soft automation, adaptive systems, and biomedical applications.

## 1. Introduction

Soft actuators, inspired by biological organisms' movement and functionalities, are designed to emulate their flexibility and adaptability [1–3]. Researchers have developed soft actuators capable of navigating challenging environments, performing delicate tasks, and interacting safely with humans [4–6]. The design of soft actuators emphasizes their ability to deform, stretch, and contract [7–10] under diverse external field actuation [11–13], e.g., light, magnetic, electric, etc. Owing to the flexibility of soft materials, these actuators can carry out intricate movements and functions without needing extra rigid components, offering significant advantages in miniaturization [14,15], particularly for applications in confined spaces and biomedical fields [16–18].

Efforts over recent decades have engendered a variety of soft actuators that exhibit rich deformations by aligning external actuation fields with advanced designs, such as modular [19–21], heterogeneous [22–24], and origami/kirigami [25,26]. Despite great advances in these designs, the actuators are typically designed to follow specific deformation modes under given external fields, forbidding their applications for complex maneuver tasks that require adaptive switching between deformation modes. Recently, self-regulated soft actuators have shown adaptive movements by employing novel actuation strategies along with unique designs [27–31]. In particular, self-regulated actuators driven by light have demonstrated remarkable oscillatory dynamics via self-shadowing effects [32,33]. However, the complex actuation method (e.g., introducing multiple laser illuminations to the actuator [34]) is typically required to achieve complex multi-bending movement or mode switching. An important next stage for intelligent soft actuators is to develop self-regulated deformations under simplified actuation conditions (e.g., a fixed single laser source).

In this work, we present Poly(*N*-isopropylacrylamide) (PNIPAM)-based soft heterogeneous hydrogel actuators which generate self-regulated deformations when stimulated by a fixed near-infrared (NIR) laser. The heterogeneous PNIPAM hydrogel actuator comprises two portions given by a pure PNIPAM hydrogel and those embedded with reduced graphene oxide (rGO). The difference in responsiveness of these two hydrogel portions forms the basis for the light-driven deformation (**Fig. 1a**). Utilizing these two building blocks (denoted as 0 for PNIPAM hydrogel and 1 for PNIPAM/rGO hydrogel) shown in **Fig. 1b-i**, we introduce three heterogeneous configurations: up-down, side-by-side and hybrid types (**Fig. 1b-ii**). These heterogeneous designs enable hydrogel actuators to achieve self-regulated multimodal deformations (**Fig. 1c**), facilitating complex movements akin to finger-bending and flexible industrial manipulators. All actuation controls are executed through a single fixed NIR laser setup, which significantly simplifies the control module and enhances operational efficiency. These advantages position the self-regulated heterogeneous hydrogel actuator as a promising candidate for future applications in soft automation, adaptive systems, and biomedical devices [35–38].

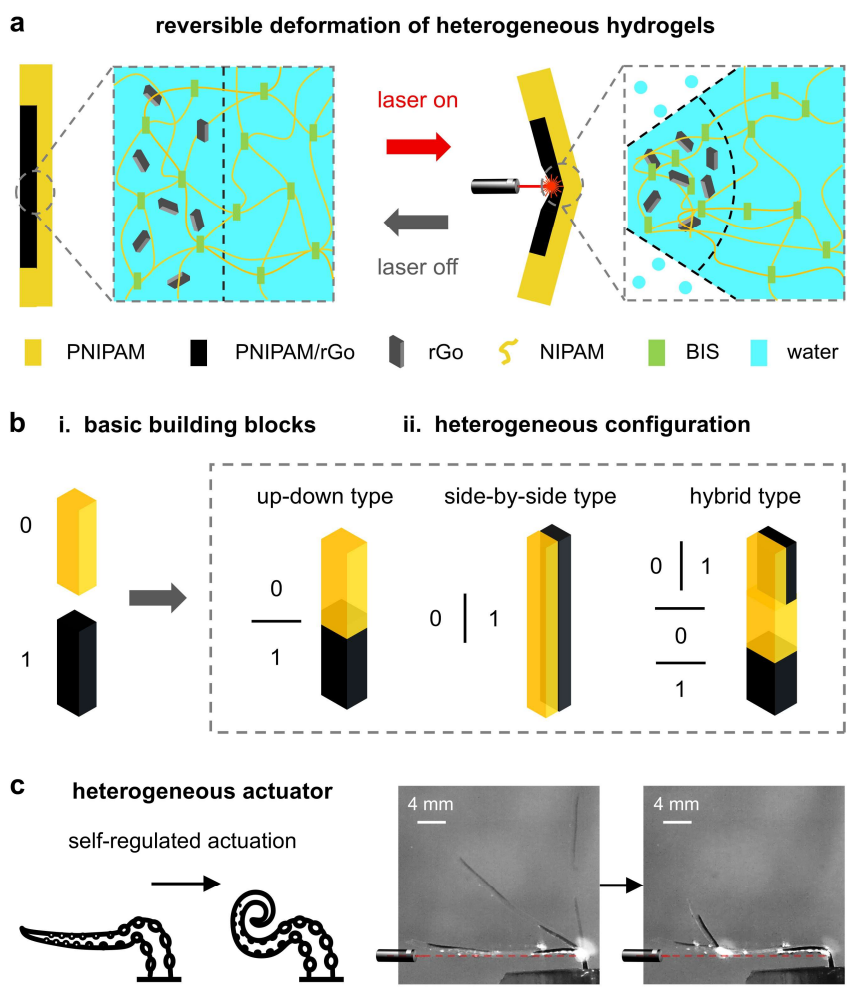

**Fig. 1** Illustration of the design concept of self-regulated heterogeneous hydrogel actuators. (a) NIR-driven reversible deformation of heterogeneous hydrogels. (b) (i) Two basic building blocks: PNIPAM hydrogel (0) and PNIPAM/rGo hydrogel (1); (ii) Three heterogeneous configurations: up-down, side-by-side and hybrid types. (c) Heterogeneous soft hydrogel actuators featuring self-regulated deformations under a fixed NIR laser.

## 2. Results and discussion

### 2.1. Fabrication and characterization of PNIPAM-based hydrogels

PNIPAM hydrogels are temperature-responsive smart materials that experience a significant phase change at their lower critical solution temperature (LCST) [39]. By incorporating photothermal agents into the hydrogels, they can respond to light stimuli, e.g., near-infrared (NIR) laser, with rapid deformation, making them ideal for applications in soft robotics, drug delivery, and tissue engineering [40–44].

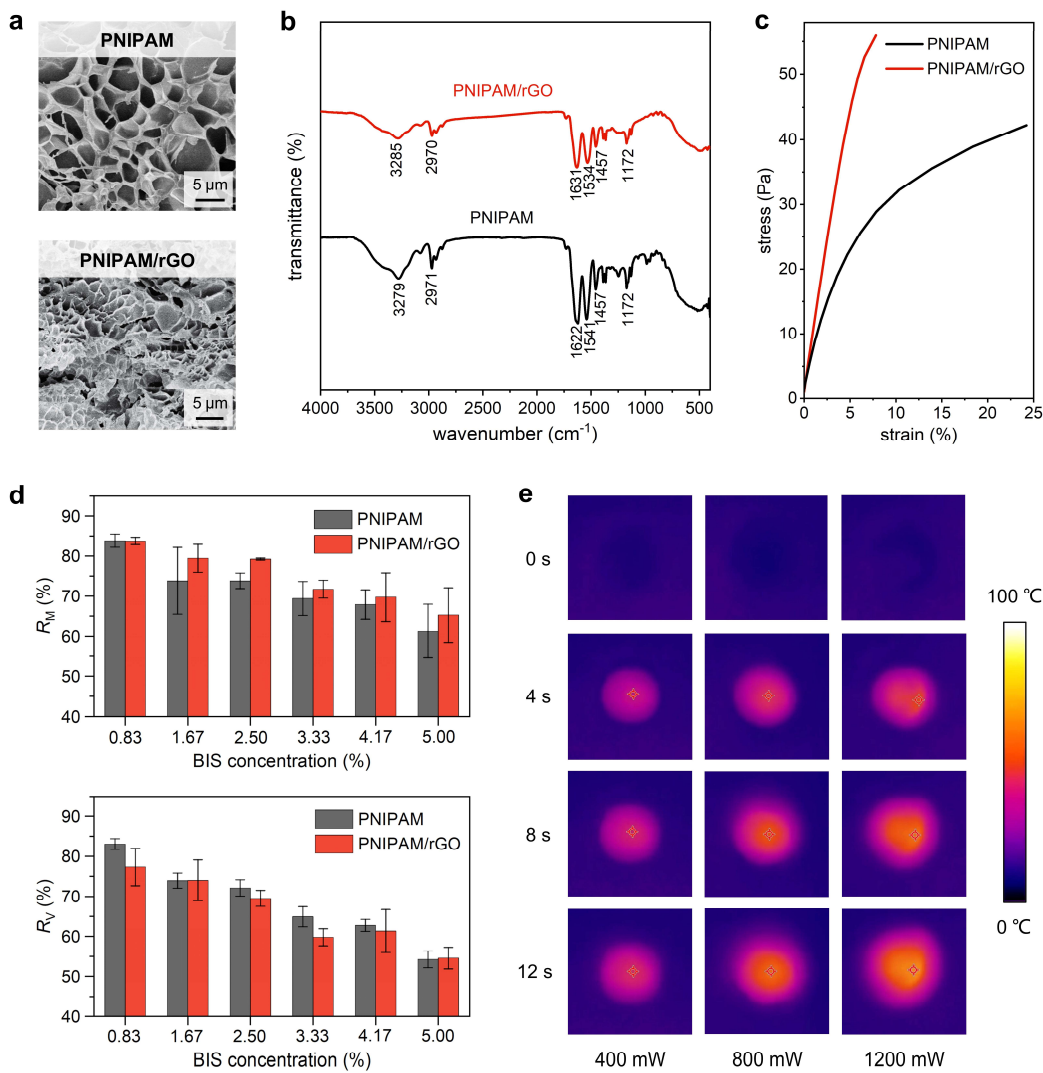

**Fig. 2** Characterization of PNIPAM-based hydrogels. (a) SEM images of PNIPAM and PNIPAM/rGO hydrogels. (b) FTIR spectra of PNIPAM and PNIPAM/rGO hydrogels. (c) Tensile stress-strain curves of PNIPAM and PNIPAM/rGO hydrogels. (d) Volume-expansion ratio and mass-change ratio of PNIPAM and PNIPAM/rGO hydrogels at different BIS concentrations (mean ± SD for 3 samples in each case). (e) Photothermal response of PNIPAM/rGO hydrogels (GO: 1.5 g/mL) under varying laser intensities (400 mW, 800 mW, and 1200 mW) over time.

Here, we introduce in-situ reduced graphene oxide (rGO) as the photothermal agent to create PNIPAM/rGO hydrogels, adding heterogeneity to the PNIPAM hydrogel. We synthesize PNIPAM and PNIPAM/GO hydrogels by curing precursors under ultraviolet (UV) light, and acquire PNIPAM/rGO hydrogels by the ascorbic acid reduction method (Experimental section and Fig. S1). Scanning electron microscopy (SEM) images show that both PNIPAM and PNIPAM/rGO hydrogels possess similar loose porous networks (**Fig. 2a**, Experimental section), which ensure their rapid response and recovery. We conduct Fourier transform infrared (FTIR) spectra of PNIPAM and PNIPAM/rGO in **Fig. 2b** (Experimental section). The characteristic peaks of PNIPAM are located at 3279 cm$^{-1}$ (secondary amide N–H stretching), 1622 cm$^{-1}$ (amide I, C=O), and 1541 cm$^{-1}$ (amide II, N–H) absorption bands [45]. In comparison, the spectrum of PNIPAM/rGO exhibits noticeable shifts to new characteristic absorption bands at approximately 3285 cm$^{-1}$, 1631 cm$^{-1}$, and 1534 cm$^{-1}$, confirming the interaction between PNIPAM and rGO in hydrogels.

**Fig. 2c** presents the tensile strength test of PNIPAM and PNIPAM/rGO hydrogels (Experimental section). PNIPAM hydrogels have inherently poor mechanical properties [46]. The doping of rGO enhances mechanical properties due to the interweaving of graphene's two-dimensional grid structure with the hydrogel during cross-linking.

In **Fig. 2d**, we evaluate the swelling properties of PNIPAM and PNIPAM-rGO hydrogels with different cross-linking agent (BIS) contents by calculating their mass-change ratio $R_M$ and volume-expansion ratio $R_V$ (Experimental section). As the concentration of BIS increases, the density of the cross-linked network also rises, resulting in a gradual decrease in the equilibrium water content of PNIPAM and PNIPAM-rGO hydrogels.

We also compare the photothermal conversion efficiency of PNIPAM/rGO hydrogel samples with varying photothermal agent contents (Experimental section and Fig. S2). We record the photothermal response of the hydrogels with optimal graphene oxide content (1.5 mg/mL) under different NIR laser intensities over time (**Fig. 2e**). The photothermal conversion rate increases with the intensity of the laser. The rapid conversion allows the PNIPAM-rGO hydrogel temperature to exceed its LCST quickly, leading to the localized contraction within heterogeneous hydrogels and thereby enabling various deformations.

The UV-curing process ensures the rapid fabrication of PNIPAM and PNIPAM/rGO hydrogels. Due to their shared base materials, these components adhere well during sequential curing, allowing for the assembly of various building blocks as shown in **Fig. 1b-ii**. We use a mold-assisted method to create hydrogels of different shapes. The low flowability of the high-viscosity PNIPAM/GO precursor (Fig. S3) supports multiple curing rounds (Experimental section). The hybrid-type heterogeneous hydrogels constructed in this manner have the potential to be developed into diverse self-regulated soft actuators.

**2.2. Self-regulated heterogeneous hydrogel actuators**

We employed a localized actuation method using a focused NIR laser as a primary actuation strategy for soft actuators [33,44], where the focused laser spot triggers rapid photothermal conversion of the hydrogels, causing localized contraction and bending deformations. We first investigate the design of an up-down type hydrogel actuator with the bottom end fixed to a PDMS base (Fig. S4). The hydrogel is immersed in water and driven by a stationary NIR laser (spot diameter: ≈3 mm), as shown in **Fig. 3a**. The diameter of the soft actuator (≈0.9 mm) is sufficient to create a difference in laser intensities between the illuminated side and the shaded side, resulting in localized contraction and bending toward the light source, even within homogeneous PNIPAM/rGO hydrogels. When the light source is turned off, the localized temperature of the hydrogel decreases below its LCST in the aquatic environment, causing water absorption and allowing the hydrogel to swell back to its original shape. We evaluate the hydrogel's ability to reversibly bend and recover via the bending angle $\theta$ (**Fig. 3a**). We consider more than one hundred switching cycles under a periodic on-off light field (**Fig. 3b-c**, Movie S1). The periodic oscillation in $\theta$ demonstrates excellent repeatability of the hydrogel actuators. Variations in laser intensity and switching frequency of the on-off light field affect the oscillatory bending behavior. At a constant frequency of 0.5 Hz (which means the laser is on for 1 s and off for 1 s, completing one full cycle every 2 s), higher laser intensity results in greater oscillation amplitudes, indicated by the increased $\theta$ (**Fig. 3b**). Conversely, at a constant laser intensity (900 mW), the median $\theta$ remains steady, but faster switching leads to smaller oscillation amplitudes (**Fig. 3c**). Additionally, we calibrate the oscillation performances of other length ratios of PNIPAM and PNIPAM/rGO hydrogels in this up-down arrangement (see details in Fig. S5).

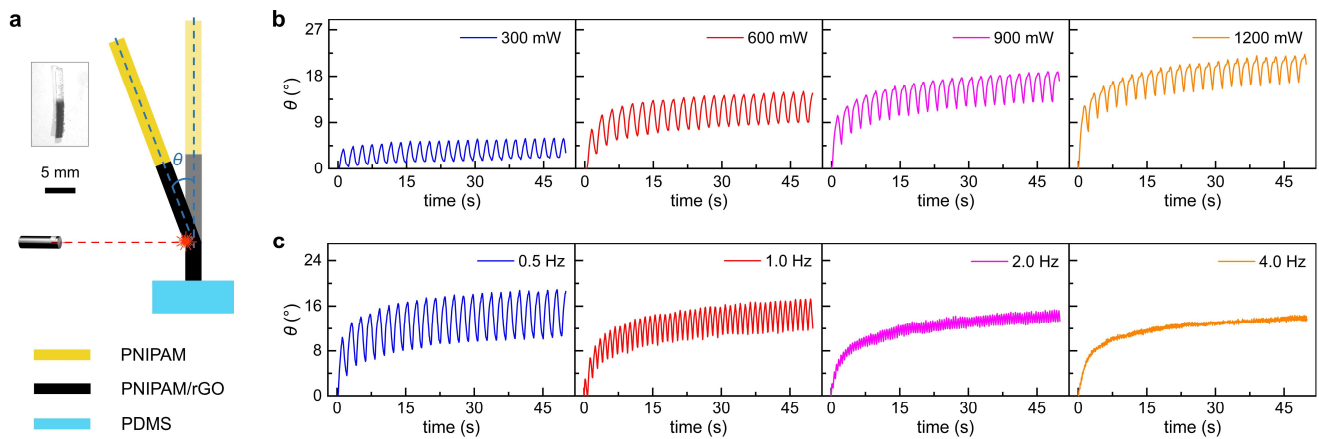

**Fig. 3** NIR laser-driven bending and oscillation of up-down type heterogeneous hydrogel actuators. (a) Illustration of the hydrogel actuator bending. (b) On-off light field (frequency: 0.5 Hz) driven actuator oscillation at different laser intensities. (c) On-off light field (light intensity: 900 mW) driven actuator oscillation at different frequencies.

Besides the above basic oscillation mode, we observe more complex, multimodal deformations under fixed and continuous laser irradiation in a hybrid heterogeneous hydrogel actuator (consisting of two side-by-side hydrogels arranged in an up-down type; Fig. S6). We observe a transition of self-regulated deformation modes with increased laser intensity $P$ (**Fig. 4**). The deformation modes are sequentially named modes 1\textendash4, based on the triggering light intensity, from low to high (Movie S2).

**Fig. 4a-i** depicts mode 1 activated at $P = 700$ mW, where the actuator exhibits a single bending deformation with a bending angle $\theta_1$. With a side-by-side configuration at the illuminated area, the exposed PNIPAM/rGO section rapidly contracts and allows the actuator to reach equilibrium at $\theta_1 \approx 82.5°$ (as shown in **Fig. 4a-ii,iii**). **Fig. 4b-i** depicts mode 2 activated at $P = 1000$ mW, where the actuator transitions to a single bending deformation with self-regulated oscillation. In this mode, the increased laser intensity causes an increase in $\theta_1$ to $\approx 85°$, which is sufficient for the actuator's top to block the irradiated area near the bottom. This eventually leads to a self-oscillating state under intermittent illumination and self-shadowing (**Fig. 4b-ii**). This self-regulation behavior is captured by the slight fluctuations in $\theta_1$ around $\approx 85°$ (**Fig. 4b-iii**). As $P$ increases to 1400 mW, the actuator transitions to mode 3, characterized by self-regulated double bending and oscillation (**Fig. 4c-i,ii**). Similar to mode 2, sufficient laser intensity triggers a self-oscillating state of the actuator. Yet, the higher laser intensity in mode 3 increases $\theta_1$ to $\approx 90°$ (**Fig. 4c-iii**). This increase in $\theta_1$ causes the PNIPAM/rGO hydrogel layer originally located on the backlight side

to move into the illuminated area during the oscillation. As a result, a secondary bending is gradually triggered from 50 s to 200 s, and reaches a stable oscillation after 200 s, as shown by the change in $\theta_2$ (**Fig. 4c-iii**). When $P$ increases to 1500 mW, the actuator transitions to mode 4, characterized by over bending (**Fig. 4d-i,ii**), and $\theta_1$ reaches ≈100º (**Fig. 4d-iii**). In this case, the upper hydrogel is no longer illuminated, and thus unable to exhibit a secondary bending. Hence, the hybrid heterogeneous configurations enable hydrogel actuators to achieve complex multimodal deformations under fixed NIR laser illumination. Our results demonstrate the intricate relationship between laser intensity and the self-regulated multimodal deformations of the hydrogel actuator.

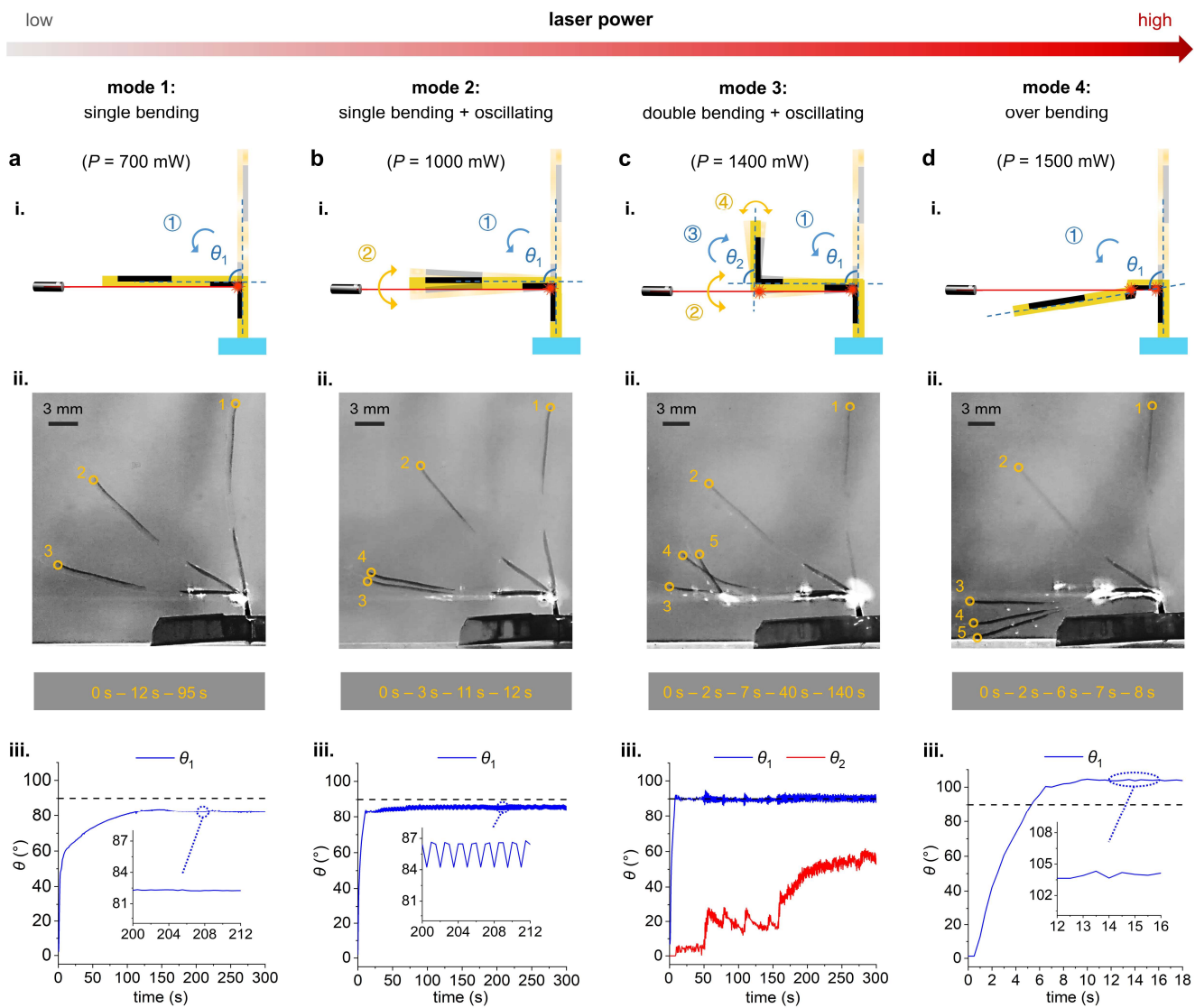

**Fig. 4** The self-regulated multimodal deformations of the hydrogel actuator with hybrid heterogeneous configurations under fixed NIR laser actuation. (a) Mode 1: single bending, including (i) illustration; (ii) three experimental snapshots superimposed based on the time sequence; (iii) variations of $\theta_1$ over time. (b) Mode 2: single bending and oscillation, including (i) illustration; (ii) four experimental snapshots superimposed based on the time sequence; (iii) variations of $\theta_1$ over time. (c) Mode 3: double bending and oscillation, including (i) illustration; (ii) five experimental snapshots superimposed based on the time sequence; (iii) variations of $\theta_1$ and $\theta_2$ over time. (d) Mode 4: over bending, including (i) illustration; (ii) five experimental snapshots superimposed based on the time sequence; (iii) variations of $\theta_1$ over time.

## 2.3. Self-regulated hydrogels for complex deformations

Inspired by the self-regulation property, we propose additional heterogeneous hydrogel designs to showcase their potential for complex deformations under fixed NIR laser actuation, including finger-like bending, and out-of-plane bending resembling the movement of flexible industrial robotic manipulators.

We propose an up-down configuration of two side-by-side arranged hydrogels as shown in **Fig. 5a** (type-1 hydrogel actuator; Fig. S7). In contrast to the arrangement shown in **Fig. 4**, here we consider PNIPAM/rGO hydrogels in the type-1 actuator positioned on the same side. This hydrogel actuator exhibits a double-bending deformation via a two-step deformation triggered by continuous illumination of two different laser intensities, where the illuminated positions function like finger joints (**Fig. 5b-c** and Movie S3). As an example to illustrate this two-step deformation, we first introduce a continuous laser at 700 mW to the hydrogels for 10 s, which initiates the first bending ① of the soft actuator. Then, we increase the laser intensity to 1100 mW to enhance the bending further, allowing light to reach the upper hydrogel section and creating a secondary bending ② there. In the motion recording, the hydrogel actuator's main axis is defined as the $Z$-axis, and the laser's illumination path is set as the $X$-axis. The motion trajectory of the actuator tip is extracted in **Fig. 5c** using the illuminated point as the origin. In **Fig. 5c**, the red and blue data points represent bending ① (0–10 s) and bending ② (10–20 s), respectively, along with arrows indicating the bending directions. The type-1 actuator undergoes two instances of clockwise bending, and exhibits increased tip trajectory curvature due to the second laser exposure point being closer to the top. This results in more complex hydrogel dynamics, leading to multi-joint bending that resembles a finger movement.

An alternative design that features up-down arranged hydrogels at the bottom and side-by-side arranged hydrogels at the top is shown in **Fig. 5d** (type-2 hydrogel actuator; Fig. S8). The illuminated homogeneous structure allows for bending in response to laser light from any direction; however, it requires a stronger actuation laser intensity to achieve the initial bending compared to the side-by-side arrangement. Therefore, we set a constant high power of $P = 1600$ mW to drive the hydrogel actuator's out-of-plane double-bending (Movie S4). During the first 10 s, the soft actuator performs bending ① in the $XZ$-plane (**Fig. 5e-f**). When the laser illuminates the upper hydrogel section, it triggers bending ② in the $XY$-plane from 10 s to 30 s (**Fig. 5e-f**). As defined by the three-dimensional coordinate system in **Fig. 5d-e**, the actuator tip's motion trajectory is extracted in **Fig. 5f** using the illuminated point as the origin. In **Fig. 5f**, the red and blue data points represent bending ① (0–10 s) and bending ② (10–30 s), respectively, along with arrows indicating the bending directions. The local change of hydrogel dynamics in the upper heterogeneous section enables the type-2 actuator to demonstrate distinct out-of-plane deformation compared to previous hydrogel deformations (**Fig. 4** and **Fig. 5a-c**), resembling flexible industrial manipulators.

The above results further demonstrate how hybrid-type heterogeneous hydrogels enable self-regulated deformations with versatile actuation modes under a single fixed NIR laser. This advancement is beneficial for developing automated soft robotics with complex dynamics. Most classical methods, whether targeting finger-like bending or in-plane/out-of-plane multi-bending deformations, often necessitate sophisticated designs and actuation mechanisms, e.g., pneumatic soft actuators with particle jamming [47], and oscillators with coordinated multiple actuation sources [34]. Here we provide an alternative approach that reduces design complexity and inspires future nature-inspired movements and adaptive systems, promising accessible and versatile robotic solutions that can dynamically respond to their environments.

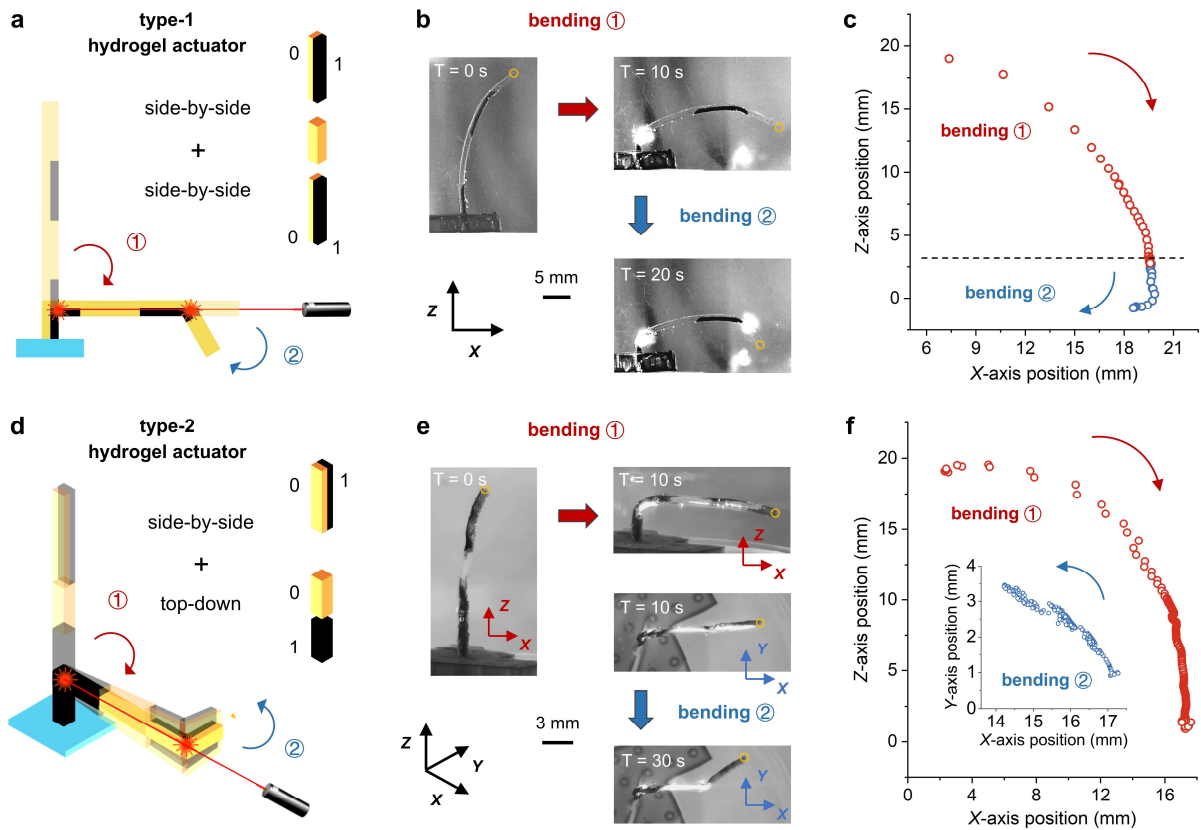

**Fig. 5** Heterogeneous hydrogel actuators for self-regulated biomimetic deformations. (a) Illustration of the heterogeneous composition of the type-1 hydrogel actuator. (b) Experimental snapshots showing the finger-like double-bending of the type-1 actuator. (c) The motion trajectory of the hydrogel actuator tip during finger-like bending. (d) Illustration of the heterogeneous composition of the type-2 hydrogel actuator. (e) Experimental snapshots depicting the out-of-plane double-bending of the type-2 actuator. (f) The motion trajectory of the hydrogel actuator tip during out-of-plane bending.

## 3. Conclusions

In this work, we present a new paradigm of self-regulated heterogeneous hydrogel actuators using a single fixed NIR laser for actuation. Through the coupling of heterogeneous designs and light field parameters to induce self-shadowing and secondary excitation effects, we achieve the self-regulation of hydrogel actuators, as verified by their multimodal deformation under varying NIR laser intensity (**Fig. 4**). Additionally, to validate the generality and scalability of this concept, we examine various types of actuators that can simulate in-plane finger-like bending and out-of-plane manipulator-like bending under the fixed NIR laser arrangement (**Fig. 5**). Future research on developing untethered actuators by using a similar heterogeneous design concept, along with alternative actuation mechanisms (e.g., magnetic fields [48,49]) or in different environments (e.g., airborne [50]), may enhance the autonomy and adaptive behavior of innovative self-regulated actuators, and ultimately transform applications in robotics, automation, and biomedical fields.

**Experimental section**

*Materials*

*N*-isopropylacrylamide (≥ 98%), *N*, *N*'-methylenebisacrylamide (BIS) (≥ 99%), 2-hydroxy-2-methylpropiophenone (≥ 97%), acrylamide (≥ 99%), and ascorbic acid (≥ 99%) were obtained from Aladdin Industrial Co., Ltd. (China). Graphene oxide aqueous solution (5 mg/ml) was provided by Tanfeng Graphene Technology Co., Ltd. (China). Sylgard 184 Polydimethylsiloxane (PDMS) and its curing agent were obtained from Dow Corning Co. (USA). Deionized (DI) water was produced by the Direct-Q deionized water system (USA). All chemicals were used as received.

*Preparation of PNIPAM and PNIPAM/rGO hydrogels*

The PNIPAM precursor was prepared by mixing *N*-isopropylacrylamide, acrylamide, BIS, 2-hydroxy-2-methylpropiophenone, and DI water in a mass ratio of 120:24:3:8:100. For the PNIPAM/GO precursor, the same mass ratio was maintained, and graphene oxide aqueous solution was added to achieve specific photothermal agent concentrations (0.5, 1, 1.5, 2.0 mg/mL). All the precursors were stirred overnight at 200 rpm, then labeled and stored away from light. PNIPAM-based hydrogels were fabricated by exposing the top and bottom sides for ≈15 s under UV light (365 nm, 300 mW/cm$^2$), respectively. In the heterogeneous design, light-curing assembly and layer-by-layer exposure were aided by molds (Fig. S4 and Figs. S6-S8). After cross-linking, the PNIPAM-GO hydrogel was immersed in a pre-prepared 0.1 g/mL ascorbic acid aqueous solution to reduce GO in the hydrogel matrix, followed by heating in a 90 °C water bath for 1 hour. The samples were then soaked in DI water (25 °C) until swelling equilibrium was reached for later use.

*Microscopy*

The SEM images were observed using a Hitachi S-4800 scanning electron microscope. Experiments were captured with a handheld microscope (Dino-Lite Edge AM4115ZTL, AnMo, Taiwan).

*FTIR spectroscopy*

The data were obtained using a Nicolet iS50 FTIR spectrometer (Thermo Scientific). The disk-shaped samples (diameter: 6 mm; thickness: 3 mm) were dried at 80 °C for 24 hours before tests.

*Tensile strength test*

The hydrogels' mechanical properties were tested using dynamic mechanical analysis (DMA) with DMA Q800 V21.2 Build 88 (TA Instruments, USA). The samples were rod-shaped hydrogels (pre-swelling dimensions: 1 mm in diameter and 20 mm in length), soaked in DI water at 25 °C for 48 hours to reach swelling equilibrium before testing.

*Swelling property characterization*

First, we measured the dry weight of the hydrogel as $M_0$ and its volume as $V_0$. Then, the hydrogel sample was placed in DI water at 25 °C for 48 hours to reach swelling equilibrium before removing them. After that, we used absorbent paper to dry the surface moisture, and recorded its weight $M_1$ and volume $V_1$. Finally, the $R_M$ and $R_V$ could be acquired using the following equations:

$$R_M = (M_1 - M_0)/M_1 \times 100\% \tag{1}$$

$$R_V = (V_1 - V_0)/V_1 \times 100\% \tag{2}$$

*Photothermal response test*

The photothermal response was induced by an 808 nm NIR laser source (TZ808AD2000-F100-GQ, Shenzhen Taizhu Technology), which was modulated by a function generator (Topward, 8150). The disk-shaped samples (diameter: 6 mm; thickness: 3 mm) were soaked in DI water (25 °C) for 48 hours until swelling equilibrium was reached, and then removed from the water for testing. The NIR laser was adjusted to focus the spot on the center of the sample, and the output power was modified on demand. Thermal images of the samples under laser irradiation (**Fig. 2e**) were captured using a thermal imager (HM-TPK20-3AQF/W, Hikvision), which recorded the temperature change over time at the same irradiation duration to determine the photothermal response performance of the hydrogels with different photothermal agent concentrations (Fig. S3).

*Viscosity measurement of the precursors*

Fig. S4 shows the viscosity of the hydrogel precursors used in this study, measured using a digital viscometer (range: 0.1–100,000 mPa·s; resolution: ±2%; SNB-2, AOYIMEI, China).

*PDMS substrate/mold preparation*

This study used PDMS molds for fabrication and PDMS substrates for hydrogel actuators. Different-shaped PDMS molds/substrates were acquired through a molding process. Specifically, PDMS and the curing agent were mixed in a mass ratio of 10:1, thoroughly stirred, and degassed for 1 hour. The mixture was then poured into the mold and heated in an oven at 70 °C for 2 hours. Finally, the sample was obtained by demolding.

*Data acquisition*

The quantitative data of deformations were extracted by using Tracker software video analysis.

**Author contributions**

J.L., J.M. and A.C.H.T. conceived this study. J.M. and A.C.H.T. supervised the project. J.L. and J.M. carried out the experiments. A.C.H.T. acquired funding. J.L., J.M. and A.C.H.T. contributed to the interpretation of the results. J.L., J.M. and A.C.H.T. drafted and revised the manuscript.

**Conflicts of interest**

There are no conflicts to declare.

**Data availability**

All data supporting the conclusions in this paper are presented in the main text and the supporting information. Additional data related to this paper may be requested from the authors.

**Acknowledgements**

The authors thank the funding support from the Croucher Foundation and equipment support from HKU Queen Mary Hospital Electron Microscope Unit.

Supporting Information for

# Self-regulated photoresponsive heterogeneous PNIPAM hydrogel actuators

Jingxuan Li[a,‡], Jiaqi Miao[a,‡], and Alan C. H. Tsang[a,*]

[a] Department of Mechanical Engineering, The University of Hong Kong, Pokfulam, Hong Kong, China
[‡] These authors contributed equally to this work.
* Correspondence: alancht@hku.hk

**This PDF file includes:**

1. Preparation of the PNIPAM-rGO hydrogel

2. Photothermal response characterization

3. Viscosity characterization

4. Calibration of the hydrogel actuator oscillation

5. Fabrication of hybrid-type heterogeneous hydrogel actuators

Figures S1-S8

Legends for Movies S1-S4

# 1. Preparation of the PNIPAM-rGO hydrogel

Fig. S1 illustrates the general fabrication method for PNIPAM/rGO hydrogels. A PDMS mold was acquired using a molding process with a 3D-printed mold (Fig. S1a). The PNIPAM/GO precursor solution was poured into the PDMS mold and cured under the UV light (Fig. S1b) as described in the Experimental section of the main text. The cured PNIPAM/GO hydrogels were reduced in an ascorbic acid solution at 90 °C for about 1 hour, leading to the formation of PNIPAM/rGO hydrogels (Fig. S1c).

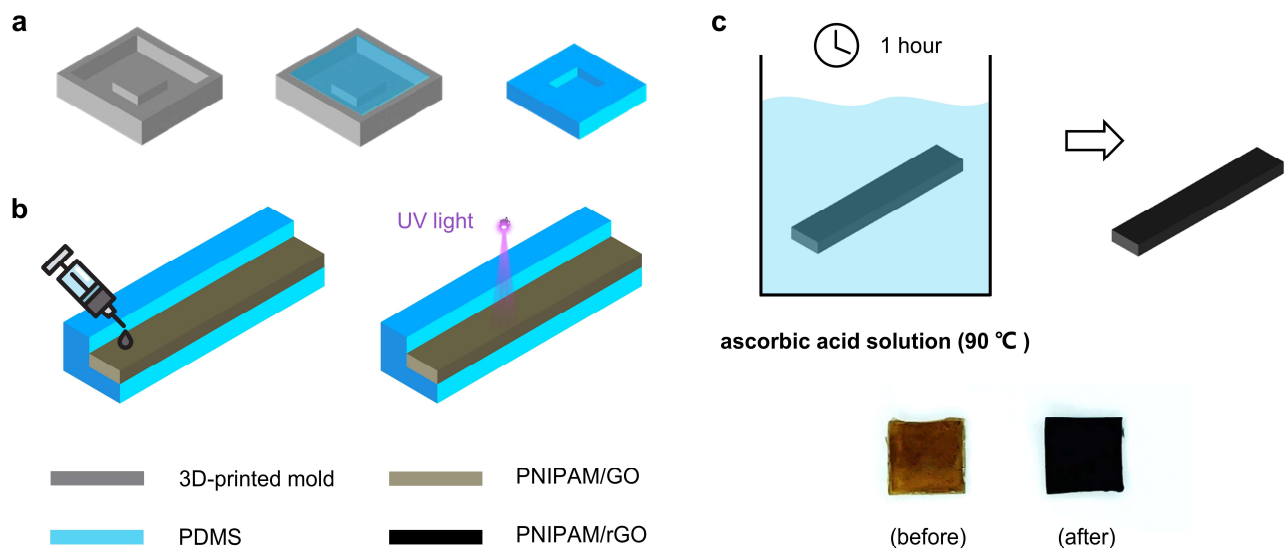

**Fig. S1.** (a) Illustration of the molding process of PDMS. (b) Illustration of the curing process of PNIPAM/GO hydrogel. (c) Reduction of the PNIPAM/GO hydrogel and acquisition of the PNIPAM/rGO hydrogel.

## 2. Photothermal response characterization

Temporal change in the highest temperature at the center of hydrogels with different GO concentrations under varying power conditions is shown in Fig. S2. Both the maximum temperature and the temperature increase rate of the hydrogels increase with higher GO concentrations. However, beyond a certain threshold, its contribution to the temperature rise rate and maximum temperature begins to diminish. The response capability peaks at a GO concentration of 1.5 mg/mL; therefore, this concentration was chosen for subsequent experiments.

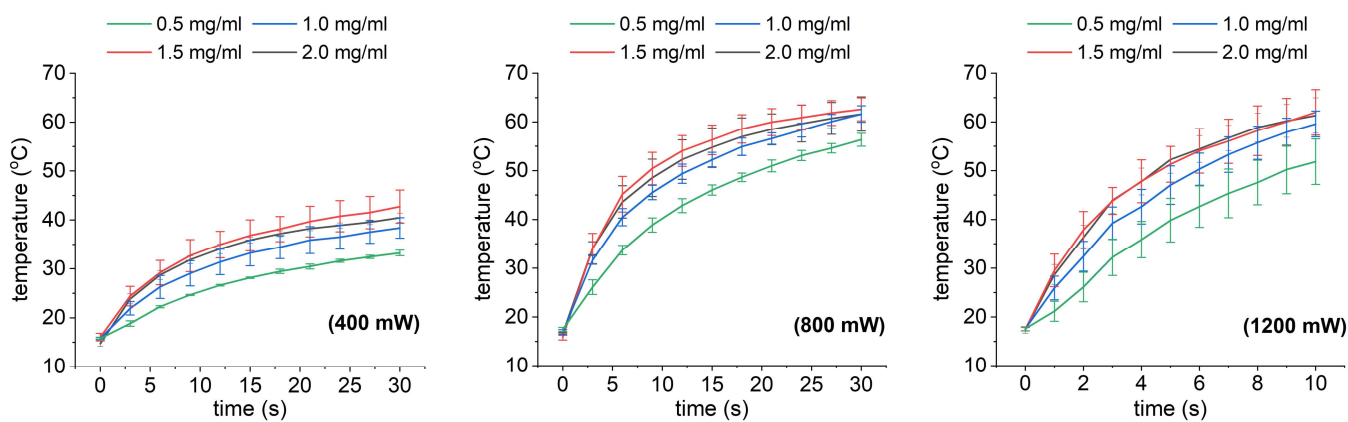

**Fig. S2.** The photothermal response curves of hydrogels with varying GO concentrations under laser intensities of 400 mW, 800 mW and 1200 mW. Error bars denote the SD.

## 3. Viscosity characterization

Viscosity measurements of the PNIPAM precursor and the PNIPAM/GO precursors at different GO concentrations, along with images of the PNIPAM precursor and the PNIPAM/GO precursor containing 1.5 mg/mL GO.

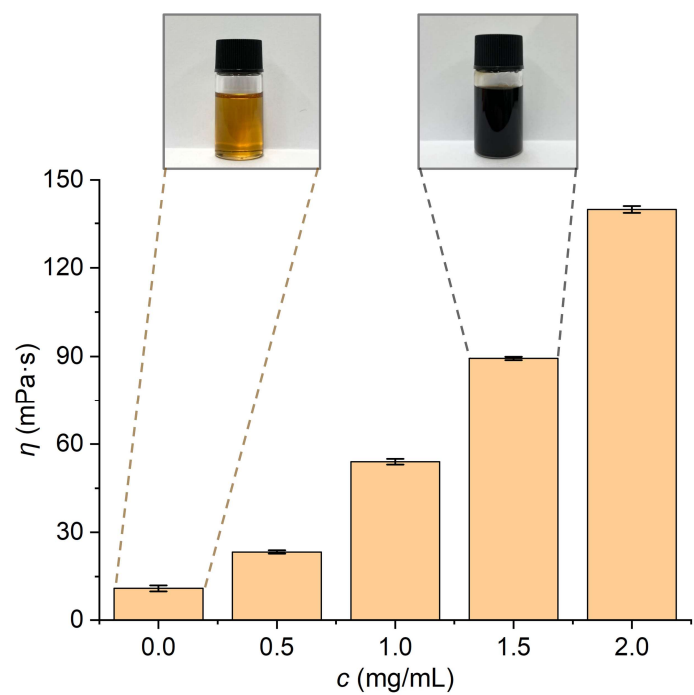

**Fig. S3.** Viscosity measurements of the PNIPAM precursor and the PNIPAM/GO precursors at different GO concentrations, along with images of the PNIPAM precursor and the PNIPAM/GO precursor containing 1.5 mg/mL GO.

## 4. Calibration of the hydrogel actuator oscillation

For the circular cross-section heterogeneous hydrogel actuator (used in Fig. 3 of the main text), the preparation process is shown in Fig. S4a. It began by cutting a syringe hose to the desired length and inner diameter, followed by a longitudinal incision along one side to allow it to open while retaining some sealing due to the hose's elasticity. A suitable volume of PNIPAM/GO hydrogel precursor solution was injected into one end of the hose and cured under UV light (300 mW/cm$^2$ for 30 s). After curing, the actuator was allowed to cool at room temperature for one minute to dissipate heat. The same volume of PNIPAM hydrogel precursor solution was then injected into the other end and cured under the same UV conditions, ensuring that the previously cured PNIPAM/GO hydrogel was masked to prevent excessive crosslinking. Once cured, the actuator was cooled for another minute before carefully peeling it from the hose along the incision. To reduce graphene oxide in the hydrogel matrix, the actuator was immersed in a 0.1 g/mL ascorbic acid solution, heated in a 90 °C water bath for 1 hour, and finally soaked in DI water at room temperature for 48 hours to achieve swelling equilibrium.

To calibrate the oscillation of heterogeneous hydrogel actuators and examine how the geometric features affect their characteristics, we fabricated different ratios (*a-b* arrangement) of PNIPAM hydrogel and PNIPAM/rGO hydrogel in the up-down type actuators (diameter: 0.9 mm; length: 10 mm), as shown in Fig. S4b. Here, *a* represents the proportion of PNIPAM/rGO hydrogel layer in the total actuator length, and *b* represents the proportion of PNIPAM hydrogel layer, with *a* + *b* = 10. A NIR laser was used to illuminate each actuator ≈3 mm from the bottom in a water environment. The experiment followed a sequence of 60 s of laser irradiation, cooling and resetting the actuator, followed by another 60 s of irradiation. The bending angle *θ* of the actuator was recorded in Fig. S5. Due to differences in structural proportions, the actuators exhibit varying response behaviors (including amplitude and frequency) even under the same actuation conditions.

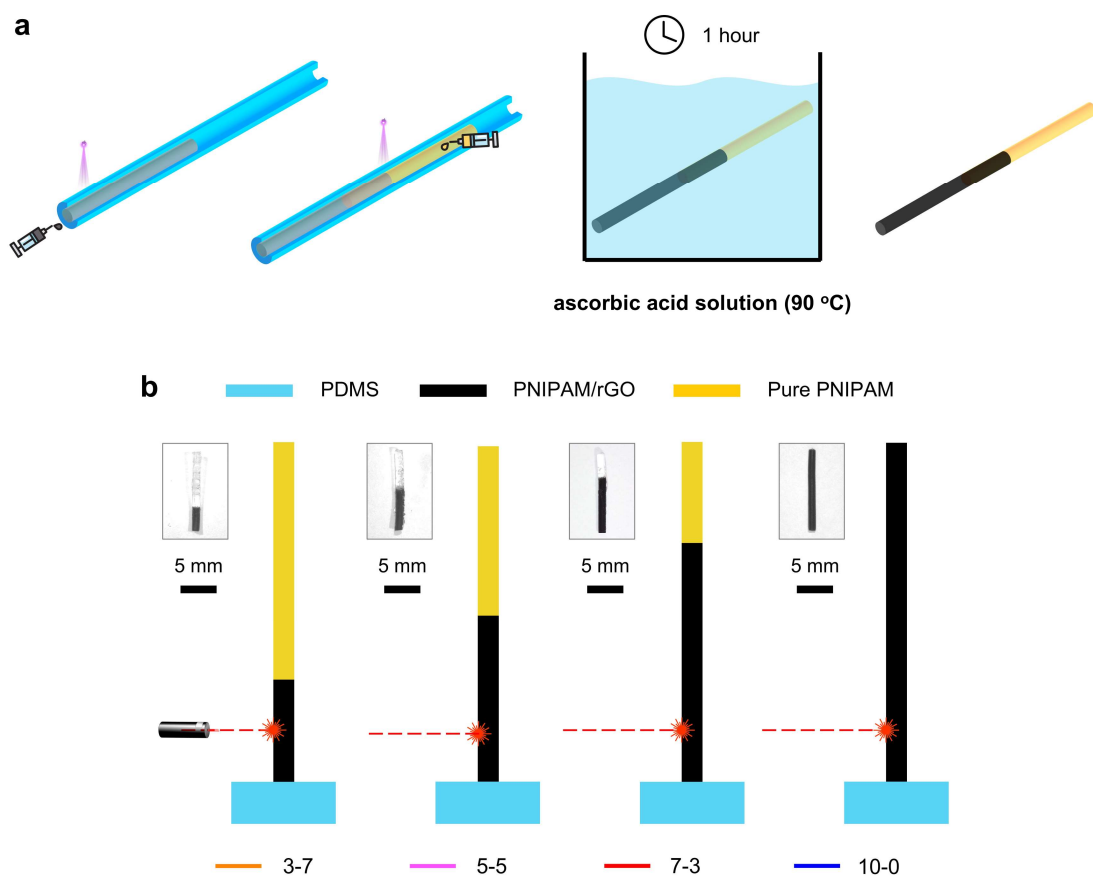

**Fig. S4.** (a) Fabrication of the circular cross-section heterogeneous hydrogel actuator. (b) Four different-ratio heterogeneous hydrogel actuators with an up-down configuration.

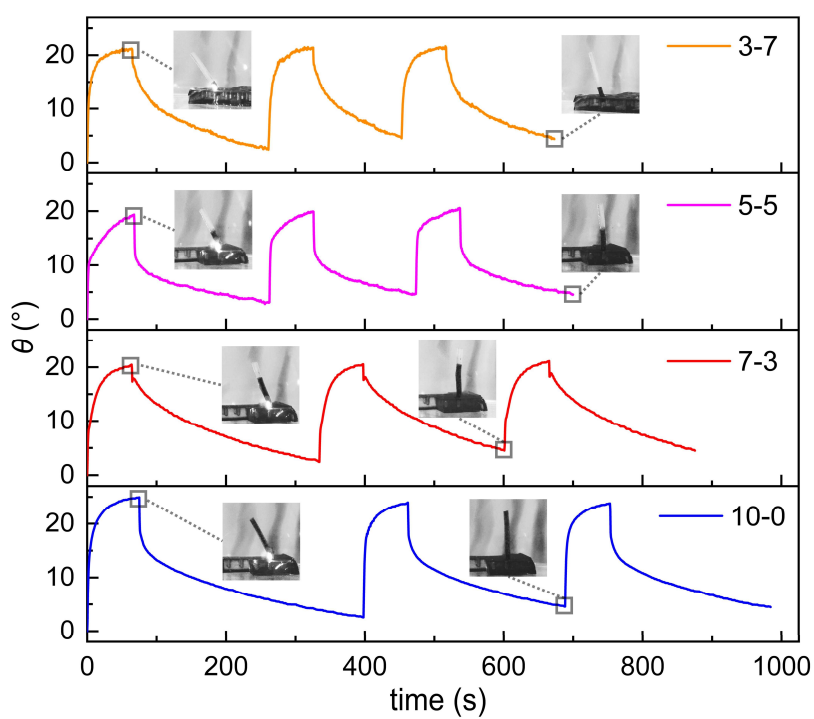

**Fig. S5.** Oscillation performance of the four types of hydrogel actuators.

## 5. Fabrication of hybrid-type heterogeneous hydrogel actuators

For the rectangular cross-section heterogeneous hydrogel actuators (cross-sectional dimensions: 1 mm × 1 mm; length: 20 mm) used in Figs. 4-5, the fabrication was performed using a molding process. First, the PNIPAM and PNIPAM/GO hydrogel precursors were degassed. Then, the high-viscosity PNIPAM/GO precursor was poured into the PDMS mold, because its viscosity allows for controlled flow and curing positions. It was exposed to UV light for about 15 s for initial curing. After cooling at 25 °C for 1 minute to dissipate heat from the cross-linking reaction, the low-viscosity PNIPAM precursor was poured into the mold to fill the remaining space. This was followed by exposure for about 15 s to complete the second curing process. The above steps could be repeated as needed to meet different structural design requirements (Figs. S6-S8). Once completed, the heterogeneous hydrogel was removed from the PDMS mold, immersed in a 0.1 g/mL ascorbic acid solution to reduce GO, heated in a 90 °C water bath for 1 hour, and finally soaked in DI water at 25 °C for 48 hours to reach swelling equilibrium.

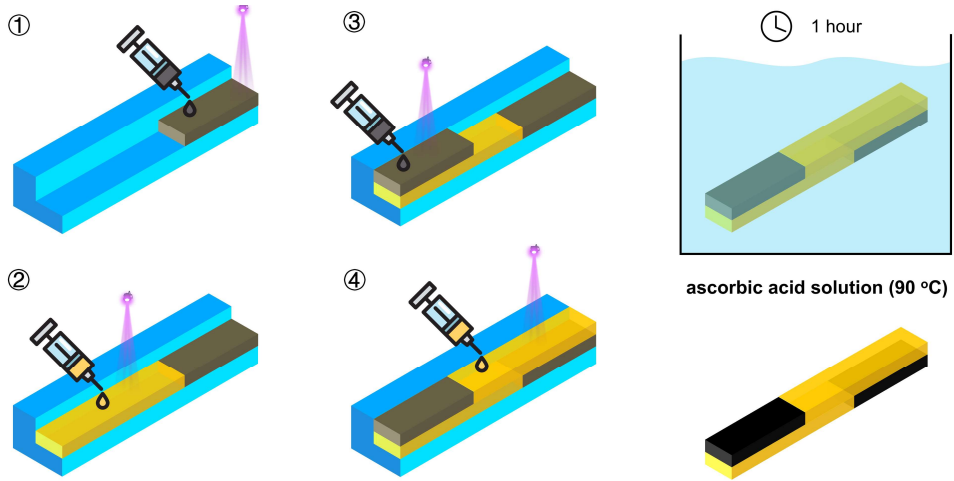

**Fig. S6.** Fabrication process of the heterogeneous hydrogel actuator used in Fig. 4 of the main text.

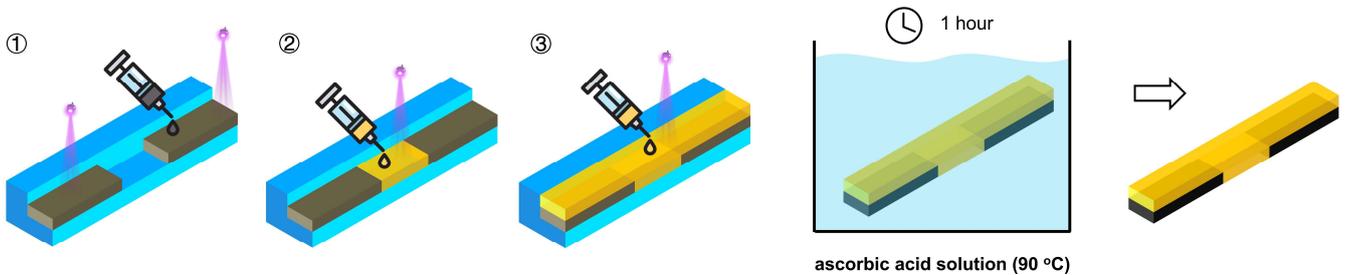

**Fig. S7.** Fabrication process of the type-1 hydrogel actuator used in Fig. 5 of the main text.

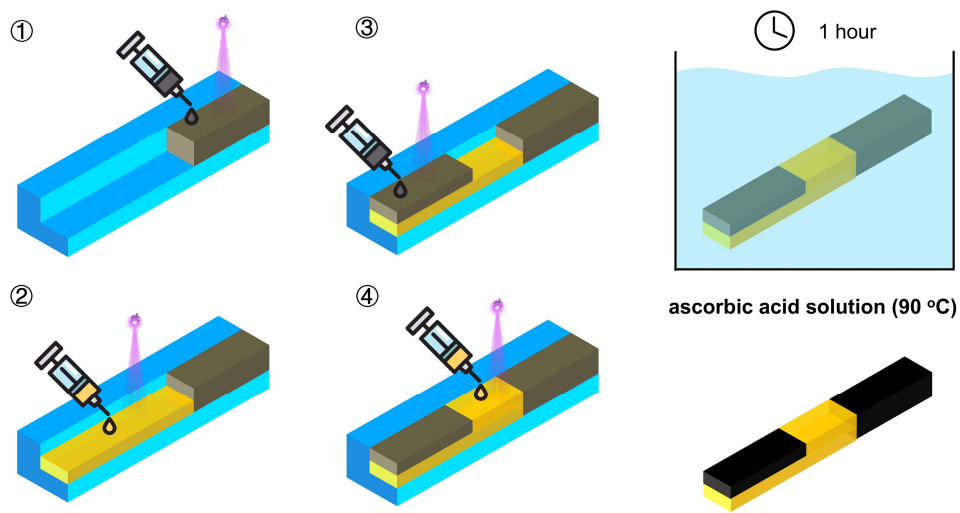

**Fig. S8.** Fabrication process of the type-2 hydrogel actuator used in Fig. 5 of the main text.